\documentclass[sigconf]{acmart}
\AtBeginDocument{%
  }

\setcopyright{acmlicensed}
\copyrightyear{2018}
\acmYear{2018}
\acmDOI{XXXXXXX.XXXXXXX}
\acmConference[CHI]{}{April 13--17,
  2026}{Barcelona, Spain}
\acmISBN{978-1-4503-XXXX-X/2018/06}

\usepackage{array}

\definecolor{darkMauve}{HTML}{B5415B}
\definecolor{darkfern}{HTML}{2B8C43}

\begin{document}

\title{Regulating AI: Where U.S. State Policy and HCI (Mis)align}

\author{Nino Migineishvili}
\email{ninom@cs.washington.edu}
\affiliation{%
  \institution{University of Washington}
  \city{Seattle}
  \state{WA}
  \country{USA}
}

\author{Alice Gao}
\email{atgao@cs.washington.edu}
\affiliation{%
  \institution{University of Washington}
  \city{Seattle}
  \state{WA}
  \country{USA}
}

\author{Adinawa Adjagbodjou}
\email{aadjagbo@andrew.cmu.edu}
\affiliation{%
  \institution{Carnegie Mellon University}
  \city{Pittsburgh}
  \state{PA}
  \country{USA}
}

\author{Dhanaraj Thakur}
\email{dthakur@cdt.org }
\affiliation{%
  \institution{Center for Democracy \& Technology}
  \city{Washington, District of Columbia}
  \state{}
  \country{USA}
}

\author{Ren\'e Just}
\email{rjust@cs.washington.edu}
\affiliation{%
  \institution{University of Washington}
  \city{Seattle}
  \state{WA}
  \country{USA}
}

\author{Katharina Reinecke}
\email{reinecke@cs.washington.edu}
\affiliation{%
  \institution{University of Washington}
  \city{Seattle}
  \state{WA}
  \country{USA}
}

\renewcommand{\shortauthors}{Migineishvili et al.}
\begin{abstract}
Artificial intelligence (AI) technologies are increasingly adopted into everyday life, with most investment and development concentrated in the U.S. In response to rapid AI integration and scant federal guidelines, U.S. states have formed AI committees charged with studying AI-related societal trade-offs. We analyzed the 18 existing state-level AI committee reports to understand how policymakers discuss AI-related benefits and risks. We then compared the risks surfaced by policymakers to an established taxonomy of AI risks aggregated from literature and examined how policymakers’ concerns align---or misalign---from those of HCI scholars. These insights provide important mileposts for shaping currently ongoing policy initiatives and future research. Our findings reveal important gaps: while committees invoke responsible AI, their framings often omit broader socio-technical concerns emphasized in HCI. We discuss opportunities for HCI to support socio-technical perspectives, employ participatory design, and close the gap between research and policy.
\end{abstract}

\ccsdesc[500]{Social and professional topics~Governmental regulations}
\ccsdesc[500]{Human-centered computing~Empirical studies in HCI}
\ccsdesc[100]{Applied computing~Law}

\keywords{Artificial Intelligence, AI, Policy, Governance, Regulation, HCI-Policy Interaction, Socio-Technical Approaches, Ethics, AI Benefits, AI Risks}
\begin{teaserfigure}
  \centering
  \includegraphics[width=\linewidth]{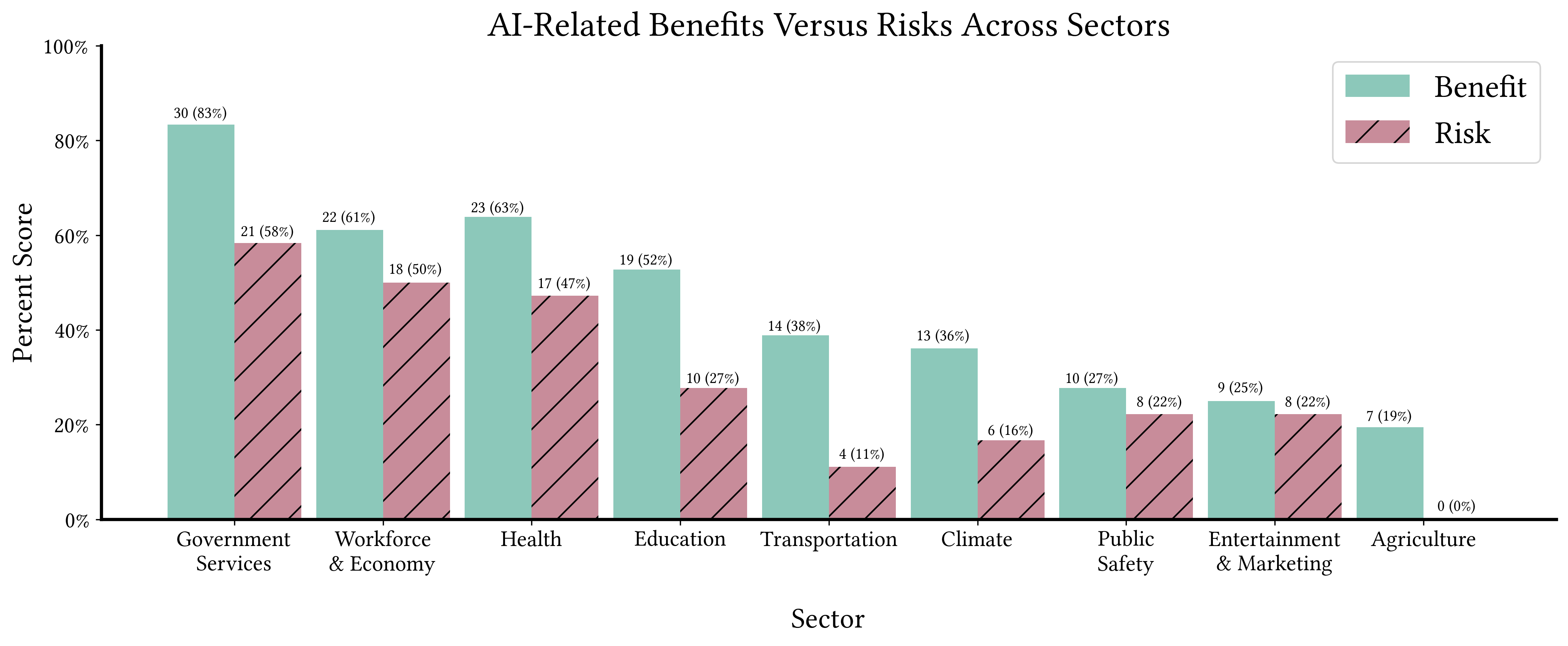}
  \caption{\textbf{AI-Related Benefits Versus Risks Across Sectors.} We assessed sectors and domains states described as benefiting from AI or being at risk from AI, scoring engagement with benefits and risks based on depth (Section \ref{sss:determining-topics}). Across sectors, states emphasize benefits more than risks.}
  \Description{This bar chart shows the sectors considered in state AI reports on the x-axis: (1) government services, (2) workforce and economy, (3) health, (4) education, (5) transportation, (6) climate, (7) public safety, (8) entertainment and marketing, and (9) agriculture. The y-axis shows the percentage of scores, calculated by summing how in-depth each sector was discussed across states. For each sector, two bars compare risks and benefits. In every case, benefits bars are higher, indicating benefits were consistently emphasized over risks.}
  \label{fig:benefit_v_risk}
\end{teaserfigure}


\maketitle

\section{Introduction}
The increasing adoption of artificial intelligence across economic sectors is leading to rapid societal shifts~\cite{abbas2024impact}, drawing the attention of governments, academia, industry, and the public~\cite{feng2025sociotechnical, schiff2020s}.
The United States is fueling much of the development, leading globally with \$100 billion in private investments in 2024 alone~\cite{maslej2025artificial}. 
This preeminence is particularly noticeable in the generative AI area, where U.S.\ private funding surpassed the combined investments of China, the United Kingdom, and the European Union~\cite{maslej2025artificial}, subsequently impacting AI practices worldwide~\cite{odeyemi2024ai}.

The scale of investment, alongside well-documented risks of AI~\cite{davenport2020artificial, wang2018deep, nordstrom2022ai}, raises an urgent need for U.S. policy to address the social implications of this rapidly evolving technology.
Surveys of the public, both in the U.S.\ and globally, consistently show that the majority of individuals are more concerned than optimistic about AI’s impact on everyday life~\cite{jackson2025google, rainie2022ai, Penn_Nesho_Ansolabehere, vesely2024survey, dreksler2025does}. 
Academics, industry professionals, and the public are therefore calling upon policymakers to define the role that AI will play in our emerging society given uncertainties around job displacement\mbox{~\cite{davenport2020artificial, Palmer_2025}}, privacy threats, or algorithmic bias~\cite{wang2018deep, horowitz2016public, bommasani2025advancing, weichert2025perceptions, O’Brien_2025}.

Yet, while global steam around AI policies intensifies---from China~\cite{Webster_Zhou_Shi_Dorwart_Costigan_Chen_2023} to the EU~\cite{act2024eu} despite lower investment---U.S. \textit{national} policy has largely stalled. U.S.\ state and local governments instead have slowly set in motion efforts to regulate AI~\cite{hatz2025local, Brennen_Perault_2023}. To this end, many states have established \textit{AI committees}, i.e., working groups tasked with studying AI, developing regulatory frameworks for it where needed, or offering policy recommendations that should govern its expansion~\cite{Brennen_Perault_2023, Engler_West_Kyooeun_Jang_MacCarthy_Mark_Muro_2024}. Published reports from these committees provide rare insights into how policymakers are actively interpreting the implications of AI and deliberating on AI-related trade-offs.

In the academic domain, Human--Computer Interaction (HCI) researchers are also  identifying and documenting AI-related benefits and risks, and shaping implementations across domains such as healthcare~\cite{reddy2020governance}, finance~\cite{lee2020access}, and the environment~\cite{kaack2022aligning}. A central concern of this scholarship is defining and shaping the ways in which AI technologies impact society and mitigating potential harms it can impose~\cite{schiff2020s, feng2025sociotechnical,nuojua2008exploring, peacock2018streets, golsteijn2015sens, thomas2017hci, henderson1998hci}.
Since policy agendas for how AI will be integrated into society are actively being set, it is critical for HCI researchers to examine what AI-related trade-offs policymakers are prioritizing and address policy proactively~\cite{thomas2017hci, yang2024future, davis2012occupy}. Still, little is known about what factors policymakers consider when legislating AI. This work accordingly poses the following guiding question: \textbf{What AI-related benefits and risks do U.S.\ state policymakers prioritize, and how do their perspectives align with those of HCI researchers?}

To answer this question, we conducted a mixed-methods analysis of the 18 existing U.S.\ state-level AI committee reports. Quantitative methods allowed us to identify high-level patterns; we employed systematic coding of motivations, benefits, and risks. We then compared the risks mentioned in committee reports with those from the AI Risk Repository~\cite{slattery2024ai}, a taxonomy of AI risks synthesized from HCI literature and closely related fields. Qualitative approaches revealed deeper nuances of how policymakers conceptualize AI; we performed a thematic analysis of mitigation strategies, values, and tensions underlying policymaking.

Our findings reveal that while state reports call for balanced, responsible AI governance and acknowledge both benefits and risks, they \textbf{systematically emphasize benefits \textit{over} risks}. Moreover, \textbf{discussions of risks tend to be cursory, lack specificity} and only partially align with those identified in research. While academic researchers stress the socio-technical nature of AI risks, committee reports emphasize isolated risks. We situate our findings within relevant policy frameworks, such as the Social Construction of Technology (SCOT) theory, and highlight HCI’s role during this active policy-setting period. Concretely, we contribute: 
(1) a typology of motivations describing why U.S.\ state policymakers consider AI, 
(2) a comparison of economic sectors policymakers see as benefiting from or at risk from AI, and a comparison of AI-related risks in state reports versus literature from HCI and related fields, 
(3) an analysis of mitigation strategies, values, and underlying tensions that shape committee recommendations, and 
(4) paths forward that emphasize broader inclusion, clearer terminology, and stronger integration between policy and HCI research.

\section{Background}

\subsection{Slow-Walking AI Policy} In recent years, governments, including those in the European Union (EU)~\cite{act2024eu}, Canada~\cite{radu2021steering} and Australia~\cite{sanderson2023ai}, have launched initiatives to regulate AI. The U.S.\ joined these efforts when the Biden Administration issued an executive order in 2023~\cite{president_2022}. The directive charged federal agencies to oversee AI development through regulation, industry guidance, and international cooperation~\cite{lubello2025biden}. 

Following the 2024 election, the Trump Administration began  dismantling these policies, framing them as barriers to innovation and issuing a new executive order to remove regulatory obstacles~\cite{prez_2025, lubello2025biden}. Most recently in November 2025, Trump signed the ``Genesis Mission'' executive order~\cite{The_White_House_2025}, allowing government agencies to share federal scientific data with private technology firms like Microsoft, OpenAI, and Google (among others) for AI-driven analysis~\cite{Gibney_Witze_Ahart_2025}.

The Trump Administration also proposed a 10-year moratorium on state-level AI regulations, threatening to withhold federal funding if states enacted rules that throttled AI expansion~\cite{Oduro_2025}. Congress ultimately rejected this proposal~\cite{Brown_O’Brien_2025}. What remained, however, was a highly unstable and volatile national-level AI policymaking landscape lacking a coherent national governance framework~\cite{davtyan2025us, parinandi2024investigating}: as \textit{The New York Times} observed, ``Washington has largely been hands-off on AI rules''~\cite{sorkin}.

\begin{figure*}[!ht]
  \centering
  \includegraphics[width=0.7\linewidth]{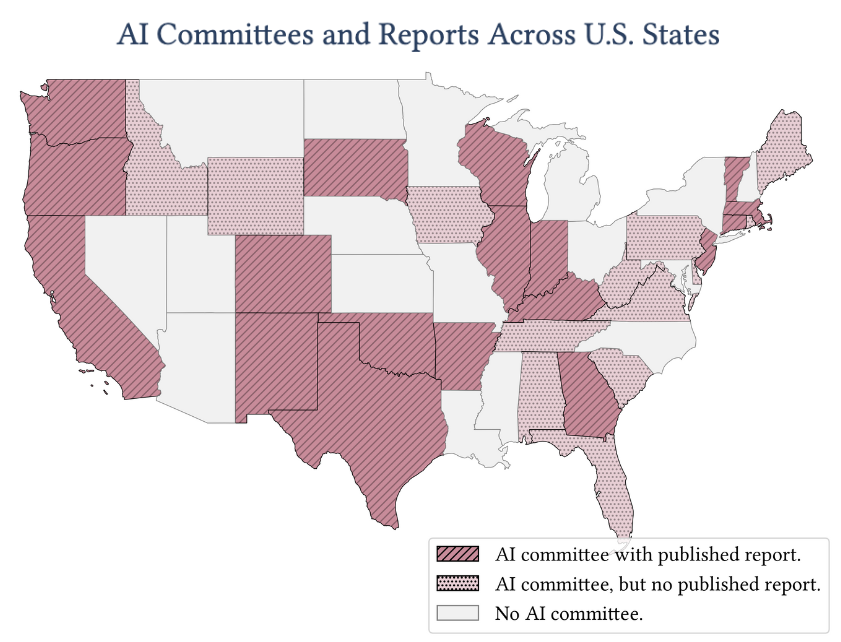}
  \caption{\textbf{AI Committees and Reports Across U.S. States:} 31 states have established AI committees (highlighted in pink, hatched), and of those, 18 have published reports (darker pink, diagonal hatches). Alaska and Hawaii did not have AI committees.}
  \Description{This figure maps U.S. states by AI committee activity: 18 states with a committee and report (dark pink), 13 with a committee but no report (light pink), and the rest with no committee (grey).}
  \label{fig:stateCommittees}
\end{figure*}

\subsection{State-Level AI Committees}
Amid this instability, many U.S.\ states have begun developing their own AI policies~\cite{davtyan2025us, parinandi2024investigating}. In 2016, only one state-level AI-related law had been enacted; by 2024, that number had surged to over 131~\cite{maslej2025artificial}. Due to limited technical expertise and institutional resources, however, policymakers often face challenges when developing AI governance strategies~\cite{hatz2025local}. In these instances, states establish committees, often called study committees, task forces, or working groups, to identify the implications of AI and propose potential regulatory frameworks~\cite{hatz2025local}. 
We refer to all such efforts collectively as \textit{AI committees} since states have no consistent or formal definition for such entities~\cite{Endersby_Ritchey_Brothers_2024}. As of this study, 31 states had entrusted committees via legislation or executive orders~\cite{Endersby_Ritchey_Brothers_2024} to explore AI-related issues in depth and pool expertise before making policy recommendations~\cite{nownes1998interest} (see Figure \ref{fig:stateCommittees}). In some cases, new committees were established to investigate AI; in others, existing committees were re-assigned to consider AI-related issues.

State-level AI committees play a key role in shaping legislation. They signal how AI policy might develop at the national level and even lay the foundation for future regulatory laws~\cite{parinandi2024investigating, hatz2025local, nordstrom2022ai}. Parinandi et al.~\cite{parinandi2024investigating}, for instance, found that states with AI committees were more likely to pass consumer protection laws. 

AI committees influence policy primarily by convening and publishing reports that summarize their findings and recommendations~\cite{sabatier2014theories}. Unlike bills, which document only the finalized outcomes of the legislative process, AI committee reports offer a window into the deliberative processes that lead to the making of a particular policy decision~\cite{schiff2020s}. They reveal the overarching concerns, areas of emerging consensus and disagreement, and the motivations shaping AI policy efforts~\cite{schiff2020s, young2019toward}. As such, committee reports offer a rich source of insight into how states are proactively approaching the challenges and opportunities of AI.

\section{Related Work}

Scholars outside of the HCI field have developed substantial frameworks for technology regulation and governance. We describe two perspectives below as they offer insights into policy priorities and perspectives current U.S.\ state policymakers might adopt when regulating AI. We then detail prior work at the intersection of HCI and AI policy.

\subsection{Perspectives on Technology Governance} 

One practice, documented by Kaminski et.~\cite{kaminski2023regulating} is that of ``\textit{ex ante risk regulation}'', which holds that policies should anticipate technology's harms and ensure that systems are safe and effective. On the surface, this appears to be a proactive, neutral approach. However, such framing can also be seen as deeply value-laden~\cite{kaminski2023regulating}. 
First, it assumes that adoption of technologies such as AI is inevitable despite harms that may accompany it~\cite{kaminski2023regulating}. 
Second, this regulatory approach may obscure certain harms, particularly those hardest to quantify, and exclude risks experienced by marginalized groups~\cite{kaminski2023regulating}.

The emphasis on benefits and risks as central categories have themselves been contested~\cite{recki2024you, luusua2020artificial}.
Researchers have challenged the assumption that AI's technical capabilities serve as reliable proxies for societal benefits~\cite{xie2025exploring}. While capabilities are tied to what AI can potentially do, their translation into benefits is fundamentally a social process. This dynamic is evident, for example, in healthcare~\cite{10.1145/3290605.3300468} and environmental management contexts~\cite{migineishvili2025wildfire}: AI decision-support systems in wildfire and forest management may help stakeholders understand wide geographies, but they offer little benefit to on-the-ground field workers already familiar with the terrain~\cite{migineishvili2025wildfire}. Within HCI, this idea is codified by the term \textit{socio-technical}. That is, social factors (e.g., stakeholders, culture, place, language, etc.) lie hand in hand with the technical capabilities of a system in inextricable ways~\cite{feng2025sociotechnical, ackerman2000intellectual, dean2021axes, ehsan2023charting}.

A second influential framework is the \textit{Social Construction of Technology (SCOT)} theory~\cite{douglas2012social, mahdavi2024social}, which derives from the social constructivist approach~\cite{basu2023three} and emphasizes that technological artifacts acquire meaning and function through interpretations by different social groups~\cite{basu2023three}. Under this view, technologies and their purposes are not fixed or inevitable but rather malleable, continuously shaped through social processes~\cite{basu2023three, woodhouse2005re}. Because SCOT is not prescriptive about which social groups should shape technologies, \textit{re}constructivist scholars instead have begun considering questions of social governance. These scholars have argued for more democratic deliberations that center marginalized voices in determining technology's meaning, function, and epistemology~\cite{woodhouse2005re, jongerden2008first}.
In our analysis, we examine the extent to which state-level committee reports echo or challenge these technology governance perspectives.

\subsection{HCI and AI Policy} 
A central concern in the HCI field is identifying and explaining how digital technologies shape human life and ensuring they serve the public good~\cite{spaa2019understanding}. Accordingly, HCI researchers have long sought to influence policy directly by testifying in state senates~\cite{lazar2015public} and by designing tools that support civic participation~\cite{manuel2020place} in urban planning~\cite{nuojua2008exploring, peacock2018streets}, environmental sustainability~\cite{golsteijn2015sens, thomas2017hci}, and health fields~\cite{talhouk2018human}.

Researchers at the intersection of HCI and AI policy have primarily focused on fairness, examining how public policy is shaped by the growing influence of algorithms used in decision-making processes~\cite{bati2018trust,
heger2022understanding}. 
Krafft et al.~\cite{krafft2021action}, for instance, frame AI as a public policy issue, supporting grassroots efforts to navigate local ordinances on AI and surveillance. 
These initiatives~\cite{Onuoha_Nucera_2022a, krafft2021action} offer important starting points for connecting HCI with AI policy, though critiques suggest that HCI’s policy engagement has often been reactive, mobilizing only after harm is evident, rather than proactively shaping the trajectory of technologies from the outset~\cite{yang2024future}. 

Therefore, as AI becomes increasingly embedded in our daily lives, HCI researchers should engage now and meaningfully with AI policy. Recent workshops have called for deeper involvement~\cite{yang2024future, feng2025sociotechnical}, highlighting policies like the EU AI Act~\cite{act2024eu} as important sources for HCI investigation. Though prior studies have examined responsible AI governance across organizations~\cite{batool2025ai}, inspected governance initiatives in Canada~\cite{attard2024governance} and researched AI ethics across public and private sectors~\cite{schiff2021ai}, \textbf{no systematic study has examined the nascent AI policy landscape in the U.S.} Importantly, this field is still crystallizing, and policy agendas are currently being made, marking an opportunity for HCI researchers to engage proactively with policy initiatives~\cite{thomas2017hci, yang2024future, davis2012occupy}.

In our study, we start to close this knowledge gap by analyzing 18 current state-level AI committee reports. These documents play a crucial role in revealing current policymaking practices, areas of focus, and key concerns. Understanding formative dynamics in the U.S.\ AI policy landscape is essential for characterizing where and how HCI can contribute to implementing technologies that empower society at large in principled and prosocial ways.

\section{Methods}
Our work examines how U.S.\ state policymakers are approaching AI governance, and it is guided by the following four research questions: 

\begin{enumerate}
\item[\textbf{RQ1:}] What motivates states to care about AI?
\item[\textbf{RQ2:}] What AI-related benefits and risks do U.S.\ state policymakers prioritize?
\item[\textbf{RQ3:}] How do the AI-related risks align with those discussed in HCI and related fields? 
\item[\textbf{RQ4:}] How do state policymakers approach mitigating AI-related risks?
\end{enumerate}

To answer these questions, we use a mixed-methods approach to analyze a set of reports issued by state-level AI committees within the legislative and executive branches.

\subsection{Positionality}
Our identities, our personal experiences, and our professional encounters influence how we approach our research~\cite{wei2024sokorsolkquantitative, fourtensions}. Thus, we describe our identities and positionality here. Our team brings varied expertise spanning computing, policy, and AI ethics. The first author spent several years working where technology and policy intersect, investigating fairness concerns in AI tools that determined whether unhoused individuals could access resources. The fourth author works at an organization dedicated to shaping technology policy, governance, and design in ways that advance equity and democratic principles. Other co-authors have extensive backgrounds in AI ethics and hold nuanced views of AI's role, recognizing areas where it can be beneficial while remaining aware of the harms it can cause in practice. We acknowledge that policymakers operate under different constraints than academic researcher and that their views and priorities might differ. However, we ground our analysis in a comparison between the views of policymakers and HCI researchers nonetheless, knowing that neither presents an ideal perspective.

\begin{table*}[!ht]
\caption{\textbf{Keywords Used To Identify State-Level AI Committees.}}
\Description{This table lists the keywords used to identify state-level AI committees. Rows group keywords by government branch, AI, and committees, which were then combined to form the final search terms.}
\begin{tabular}{@{} l p{12cm} @{}}
\toprule
\textbf{Category} &\textbf{Keyword}
\\ \midrule
Branch & ``legislature'' or ``general assembly'' or ``legislative assembly'' or ``governor'' or ``senate'' or ``house of representatives'' or ``house of delegates'' or ``assembly''\\ \midrule

AI & ``AI'' or ``artificial intelligence'' \\ \midrule

Committee & ``committee'' or ``working group'' or ``advisory council'' or ``task force'' \\ \bottomrule
\end{tabular}
\label{tab:search}
\end{table*}

\subsection{Definitions and Scope}
Our focus is on understanding how policymakers identify and prioritize AI-related benefits and risks as they \textit{actively} consider future policy interventions. This led us to adopt the following definitions and scope:

\paragraph{AI.} We intentionally adopt a broad definition of AI. Although we are interested in policy and responses to advances in large generative AI models (LGAIMs)~\cite{hacker2023regulating} such as ChatGPT or Stable Diffusion, regulatory frameworks, including the EU AI Act, tend to address both conventional AI systems and LGAIMs~\cite{hacker2023regulating}. To preserve this broader perspective, we keep our definition of AI flexible and consider both traditional and emerging AI as within scope of this study.

\paragraph{AI Policy.}
There are no standardized definitions of AI public policy~\cite{gairola2025public} or AI governance~\cite{batool2025ai} in HCI. Instead, we define AI policies as \textit{laws or regulations at the federal, state, or local level aimed at guiding how AI is integrated into society}, adapted from Krafft et al.~\cite{krafft2021action}. Further, we focus on AI policy solely in the U.S.\ context.

\paragraph{State-Level AI Committees.} 
In the absence of comprehensive federal AI regulations, state governments are beginning to develop their own guidelines~\cite{fpf}. Accordingly, our study focuses only on state-level AI committees, which offer insights into how states are actively defining and addressing AI policy and regulation. Moreover, we confine our search to AI committees formed only within state legislative or executive branches. We exclude those formed by the judicial branch since they tend to focus on how existing policies should be interpreted or enforced \textit{after} laws, policies and regulations have already been determined, not before.

\subsection{Identifying AI Committees and Reports}
To identify which U.S. states had formed an AI committee, we constructed search queries using combinations of three keyword categories: (1) the relevant governmental branch, (2) terms referring to artificial intelligence, and (3) terms denoting committees. For the first category, we limited our scope to state legislative and executive bodies, using keywords such as ``state legislature'' or ``state governor.'' However, there is variation in state nomenclature, with some states referring to their legislature as ``general assembly'' or ``legislative assembly.'' We included those as keywords to ensure full coverage. For the second category, we include two variations of the term AI, namely ``artificial intelligence'' and ``AI.'' For the third category, we use keywords such as ``task force'' and ``working group'' to capture variations in committee titles. This approach  strikes a balance between relevance and coverage. Table \ref{tab:search} summarizes the full sets of keywords we used in our searches.

We generated search strings by taking the cross product of all keyword combinations, resulting in 64 distinct queries. We applied the queries to each of the 50 U.S.\ states. For example, to locate the AI committee in Virginia, we tested queries such as ``Virginia state legislature AI committee'' and ``Virginia house of delegates artificial intelligence advisory council.'' Each search term was entered into Google, and the top 10 search results were manually inspected for relevance. A search result was deemed relevant if it led to the website of the AI committee and was formed by the state legislative or executive branch. Alternatively, the result was also deemed relevant if it led to the website of a legislative bill or executive order that proposed the creation of the AI committee and had passed. For every state, we tested each of the 64 queries until we identified a relevant committee. If none was found after all search strings had been exhausted, we determined that the state had no AI committee. In total, we identified 31 states (62\%) that had established an AI committee (Figure~\ref{fig:stateCommittees}).

AI committees often produce summary reports, typically on an annual basis throughout their tenure. These reports provide insight into how policymakers are interpreting AI-related issues and are central to our research. For each identified AI committee, we investigated whether it had published a report by reviewing the committee's  website. We documented whether a report and a website existed.  In total, we found that 18 (36\%) states had released at least one report. Table \ref{tab:reportSummary} lists these states.

\begin{table*}[!ht]
\caption{\textbf{States with AI Committees and Published Reports.}}
\Description{The table shows the 18 states with AI committees that have published reports. For each state, Arkansas, California, Colorado, Connecticut, Georgia, Illinois, Indiana, Kentucky, Massachusetts, New Jersey, New Mexico, Oklahoma, Oregon, South Dakota, Texas, Vermont, Washington, and Wisconsin, it lists the committee name, year of formation, report title, and publication year.}
  \label{tab:reportSummary}
  \begin{tabular}{@{} p{0.05\linewidth} 
  p{0.30\linewidth} 
  p{0.04\linewidth}
  p{0.30\linewidth} 
  p{0.04\linewidth} @{}}
    \toprule
      \textbf{State} &
      \multicolumn{2}{c}{\textbf{Committee}} &
      \multicolumn{2}{c}{\textbf{Report}} \\
      \cmidrule(lr){2-3} \cmidrule(lr){4-5}
       & {Title} & {Year} & {Title} & {Year} \\
      \midrule
     
     \ 1. AK & \href{https://www.transform.ar.gov/wp-content/uploads/Arkansas-AI-Center-of-Excellence_2024.pdf}{The Arkansas AI and Analytics Center of Excellence (AI CoE) Working Group} & 2024 & \href{https://governor.arkansas.gov/arkansas-ai-and-analytics-center-of-excellence-initial-report/}{Initial Report: Arkansas Artificial Intelligence and Analytics Center of Excellence} & 2025 \\ \midrule

     \ 2. CA &   \href{https://www.govops.ca.gov/generative-ai-genai-executive-order/}{Generative AI (GenAI) Executive Order} & 2023 & \href{https://www.govops.ca.gov/wp-content/uploads/sites/11/2023/11/GenAI-EO-1-Report_FINAL.pdf}{State of California Benefits and Risks of Generative Artificial Intelligence Report} & 2023 \\ \midrule

     \ 3. CO &   \href{https://leg.colorado.gov/committees/artificial-intelligence-impact-task-force/2024-regular-session}{Artificial Intelligence Impact Task Force} & 2024 & \href{https://leg.colorado.gov/sites/default/files/images/report_and_recommendations_0.pdf}{Report and Recommendations} & 2025 \\ \midrule

     \ 4. CT &   \href{https://www.cga.ct.gov/gl/taskforce.asp?TF=20230720_Task\%20Force\%20to\%20study\%20A.I.,\%20and\%20develop\%20an\%20A.I.\%20bill\%20of\%20rights}{Connecticut Artificial Intelligence Working Group} & 2023 & \href{https://www.cga.ct.gov/gl/tfs/20230720_Task\%20Force\%20to\%20study\%20A.I.,\%20and\%20develop\%20an\%20A.I.\%20bill\%20of\%20rights/20240201/CT\%20AI\%20Working\%20Group\%20Report.pdf}{CT AI Working Group Report} & 2024 \\ \midrule

     \ 5. GA &   \href{https://www.legis.ga.gov/legislation/66281}{Georgia Senate Study Committee on Artificial Intelligence} & 2024 & \href{https://www.legis.ga.gov/api/document/docs/default-source/senate-press-office-document-library/2024/study-committees-2024/ai-final/ai-report-final-signed-no-appendix.pdf}{Final Report of the Senate Study Committee on Artificial Intelligence} & 2024 \\ \midrule

     \ 6. IL &    \href{https://govappointments.illinois.gov/boardsandcommissions/details/?id=a6cedb70-8a2d-ee11-9965-001dd809cbbc}{Generative AI and Natural Language Processing Task Force} & 2023 & \href{https://www.ilga.gov/reports/ReportsSubmitted/5432RSGAEmail11660RSGAAttachAI\%20Task\%20Force\%20Report\%20final.pdf}{Report of the Generative AI and Natural Language Processing Task Force} & 2024 \\ \midrule

     \ 7. IN &   \href{https://iga.in.gov/2024/committees/interim/artificial-intelligence-task-force}{Artificial Intelligence Task Force} & 2024 & \href{https://iga.in.gov/publications/committee_report/ai-task-force-final-report-2024.pdf}{Indiana Artificial Intelligence Task Force Final Report} & 2024 \\ \midrule

     \ 8. KT &   \href{https://legislature.ky.gov/Committees/Pages/Committee-Details.aspx?CommitteeRSN=424&CommitteeType=Special+Committee}{Special Committee Artificial Intelligence Task Force} & 2024 & \href{https://apps.legislature.ky.gov/CommitteeDocuments/383/31018/11\%2013\%202024\%20AI\%20Task\%20Force\%20Findings\%20and\%20Recommendations\%20Memo.pdf}{Findings and Recommendations of the Artificial Intelligence Task Force} & 2024 \\ \midrule

     \ 9. MA &   \href{https://boards.mass.gov/detail/100180/ai-strategic-task-force}{AI Strategic Task Force} & 2024 & \href{https://www.mass.gov/doc/massachusetts-ai-strategic-task-force-2024-report-to-the-governor/download}{Massachusetts AI Strategic Task Force} & 2024 \\ \midrule

     10. NJ &   \href{https://innovation.nj.gov/projects/ai-task-force/}{Artificial Intelligence Task Force} & 2023 & \href{https://innovation.nj.gov/news/NJ-AI-Task-Force-Report.pdf}{Report to the Governor on Artificial Intelligence} & 2024 \\ \midrule

     11. NM &   \href{https://www.nmlegis.gov/committee/Interim_Committee?CommitteeCode=STTC}{Science, Technology \& Telecommunications Committee} & 2023 & \href{https://www.nmlegis.gov/Publications/Interim_Reports/STTC23.pdf}{Interim Final Report} & 2023 \\ \midrule

     12. OK &   \href{https://oklahoma.gov/governor/newsroom/newsroom/2023/september2023/governor-stitt-announces-ai-task-force.html}{Oklahoma AI Task Force} & 2023 & \href{https://oklahoma.gov/content/dam/ok/en/governor/documents/Task\%20Force\%20Emerging\%20Technologies\%20AI\%20Strategy\%20for\%20State\%20Agencies\%20in\%20OK.pdf}{Artificial Intelligence Strategy to Support State Agencies in Oklahoma} & 2023 \\ \midrule

     13. OR &   \href{https://apps.oregonlegislature.gov/liz/2023I1/Committees/JTFAI/Overview}{Oregon Task Force on Artificial Intelligence} & 2024 & \href{https://olis.oregonlegislature.gov/liz/2023I1/Downloads/CommitteeMeetingDocument/287623}{Final Report and Recommendations Joint Task Force on Artificial Intelligence} & 2024 \\ \midrule

     14. SD &   \href{https://sdlegislature.gov/Interim/Committee/483/Detail}{Study Committee on Artificial Intelligence and Regulation of Internet Access by Minors} & 2024 & \href{https://mylrc.sdlegislature.gov/api/Documents/Attachment/269419.pdf?Year=2024}{Final Report} & 2024 \\ \midrule

     15. TX &   \href{https://aiadvisorycouncil.texas.gov/s/}{Texas Artificial Intelligence Advisory Council} & 2023 & \href{https://www.house.texas.gov/pdfs/committees/reports/interim/88interim/House-Select-Committee-on-Artificial-Intelligence-and-Emerging-Technologies-Interim-Report-2024.pdf}{Final Interim Report} & 2024 \\ \midrule

     16. VT &   \href{https://accd.vermont.gov/economic-development/artificial-intelligence-task-force/}{Vermont Artificial Intelligence Task Force} & 2018 & \href{https://outside.vermont.gov/agency/ACCD/ACCD_Web_Docs/ED/MajorInitiatves/ArtificialIntelligenceTaskForce/FinalReport.pdf}{Artificial Intelligence Task Force Final Report} & 2020 \\ \midrule

      17. WA & \href{https://www.atg.wa.gov/aitaskforce}{Washington State Artificial Intelligence Task Force} & 2024 & \href{https://agportal-s3bucket.s3.us-west-2.amazonaws.com/uploadedfiles/2024\%20AI\%20Task\%20Force\%20Report.pdf?VersionId=2hKCaDeJaZ5L6BtFX9oqQg_WaY.KOCpw}{Inaugural Report of the Washington State Artificial Intelligence Task Force} & 2024 \\ \midrule

      18. WI & \href{https://dwd.wisconsin.gov/ai-taskforce/}{Governor's Task Force on Workforce and Artificial Intelligence} & 2023 & \href{https://dwd.wisconsin.gov/ai-taskforce/pdf/ai-advisory-action-plan.pdf}{Advisory Action Plan} & 2024 \\ 
      
      \bottomrule 
  \end{tabular}
\end{table*}

\subsection{Data Analysis}
Since committee reports varied in structure and we aimed to identify commonalities and differences, we employed a mixed-methods approach. We analyzed RQ1--3 quantitatively and RQ4 qualitatively. The quantitative analysis required deductive (Section \ref{sss:method-codebook}) and systematic (Sections \ref{sss: method-motivation}--\ref{sss:determining-risks}) coding of policy reports, while the qualitative analysis relied on inductive coding (Section \ref{sss:method-codebook}) and thematic analysis (Section \ref{sss:thematic-analysis}). All coding processes were iterative, collaborative, and open~\cite{thomas2006general}.

\subsubsection{Generating the Codebook.}\label{sss:method-codebook}
To generate a codebook, seven AI committee reports (39\% of the 18 total) were randomly sampled. One report was assigned to three researchers, and the remaining six were each reviewed by two researchers. We anticipated variation in report scope and focus; thus, this approach ensured that all researchers had one report that served as a common reference point for calibration. However, having diversity in the remaining reports gave us more coverage of potential themes in the reports. Each researcher closely read their assigned reports and independently generated preliminary codes, both through deductive and inductive coding simultaneously~\cite{fereday2006demonstrating}.  
\begin{itemize}
    \item \textbf{Deductive Coding.} We were interested in motivations as well as AI-related benefits and risks committees prioritized (RQ1--3). Thus, researchers created codes for motivations, focus areas perceived to be enhanceable by AI, and AI-related risks cited in the reports.
    \item \textbf{Inductive Coding.} We followed an inductive thematic analysis approach~\cite{braun2006using} to identify underlying patterns in committee recommended mitigation strategies (RQ4).
\end{itemize}

The team then met to compare and refine the codes, resolving discrepancies and updating the code set, as needed. We added and removed codes and established shared definitions to ensure consistency. 
Finally, one author applied the finished codes to all 18 reports and findings were presented to the remaining authors for agreement. The finalized codebook can be found in Appendix \ref{appendix:codebook}.

\subsubsection{Motivations Behind Forming Committees (RQ1).}\label{sss: method-motivation}
We aimed to identify the underlying motivations behind policymakers’ interest in AI. Using data from the deductive coding process (Section \ref{sss:method-codebook}), three researchers developed a typology of motivations from the reports. A motivation was considered valid if it was explicitly stated as a committee mission or goal. Disentangling the various factors driving the creation of AI committees within state governments was challenging. State motivations were rarely singular or clearly articulated. Instead, states layered multiple rationales into hybrid justifications, making it difficult to distinguish primary drivers from secondary ones. Thus, we report on the predominant motivations mentioned---economic growth, government operations, and responsible governance---but cannot assess their relative importance to one another.

\subsubsection{AI-Related Trade-Offs (RQ2).}\label{sss:determining-topics}
We examined how policymakers characterized AI’s benefits and risks, focusing on sectors they saw as either gaining from AI or facing new challenges. To highlight areas of policy relevance and priority, three researchers used codes from the deductive coding process  (Section \ref{sss:method-codebook}) to develop a typology of domains mentioned in the reports. A domain was included if reports cited specific examples of AI’s potential to optimize or support tasks, or if they identified risks or harms in that specific area. This process yielded nine common domains.

To quantify how benefits and risks were prioritized for each topic, we applied the systematic scoring procedure of Schiff et al.~\cite{schiff2021ai}. In particular, for each state, we considered how benefits and risks for every topic were discussed. We assigned a score of zero, one, or two to benefits for a topic per these  specifications: (1) A score of zero indicated benefits (risks) for topic were absent. (2) A score of one indicated benefits (risks) for that topic were mentioned only briefly or in passing but without sufficient development or substantive discussion. (3) A score of two indicated benefits (risks) for that topic was addressed substantively. We considered a topic to be substantive if it was the focus of an entire section or was developed using a definition, multiple examples, or recommendations.

Reports varied in length from 5 to 100 pages. To avoid penalizing shorter reports, we coded topics as substantial if our reading made it clear the topic had been a key consideration. Similarly, committee reports differed in style. Some consolidated committee discussions, and others summarized content from external presentations. In 
the latter case, we classified reports as 1 if a single panelist had mentioned benefits (risks) of that topic in passing, and 2 if either multiple panelists had discussed benefits (risks) of that topic or one panelist had discussed benefits (risks) of that topic in depth.

\subsubsection{Comparison with Literature in HCI (RQ3).}\label{sss:determining-risks}
Our goal here was to quantitatively assess how often AI risks appeared in committee reports versus in HCI and related field research. To ensure a fair comparison, we had two desiderata:
\begin{itemize}
    \item \textbf{D1.} The definitions of AI in the literature had to be as broad as in our analysis. A focus only on LGAIMs or only on traditional AI would not enable a balanced comparison given the broad and often undefined scope of AI in committee reports.
    \item \textbf{D2.} The literature had to address AI risks in a structured, categorized way. Without such a taxonomy, it would be impossible to build a shared understanding or to comprehensively classify risks.
\end{itemize}

We identified the AI Risk Repository~\cite{slattery2024ai} as best meeting these criteria. This repository synthesizes over 700 risks identified from 65 documents from literature in HCI and related fields into a taxonomy of AI-related risks, making it both comprehensive and widely used as a reference. It fulfills D1 by adopting a broad definition of AI that includes both traditional and generative models~\cite{slattery2024ai}, thus enabling fair prevalence comparisons; it fulfills D2 by providing a categorical treatment of risks. 

Though HCI is not the explicit home discipline of the AI Risk Repository, we believe it to nevertheless characterize AI discourse in HCI and related fields. It pulls from AI ethics~\cite{shelby2023sociotechnical, sherman2024ai, saghiri2022survey, ferrara2024genai}, fairness and accountability~\cite{weidinger2022taxonomy}, and technology governance~\cite{teixeira2022exploratory} scholarship. These disciplines address overlapping questions about AI's societal impacts, and capture the interdisciplinary ways that HCI researchers employ when studying AI.

Further, we acknowledge that our approach contains an inherent asymmetry. We are comparing committee reports (which vary in whether and how extensively they discuss risks) against research literature specifically selected for its focus on categorizing AI risks. This comparison, while asymmetric, reflects our research question to understand whether and how the risks committee reports do discuss align with those of HCI researchers.




We utilized the March 2025 version of the repository, the most recent as of our study. The repository classifies risks into seven domains: (1) Discrimination \& toxicity, (2) Privacy \& security, (3) Misinformation, (4) Malicious actors \& misuse, (5) Human-Computer Interaction, (6) Socioeconomic \& environmental, and (7) AI system safety, failures, \& limitations, which are further subdivided into a total of 24 risk categories. Many risks span multiple domains~\cite{slattery2024ai}. To align committee report risks with this taxonomy, we adopted the procedure used in the repository. We first coded any explicit mention of a risk, challenge, or concern as ``risk.'' For instance, we did not code ``data privacy is important'' but did code ``there is a risk that individual data is leaked.''

We then mapped each risk from the reports to one or more subdomains in the AI Risk Repository, consistent with the repository’s definitions. Multiple assignments were possible due to its lack of mutual exclusivity. For example, ``AI models trained on biased data can perpetuate societal biases, leading to unequal outcomes'' was mapped to both ``Unfair discrimination and misrepresentation'' and to ``Unequal performance across groups.'' One researcher performed the matching, discussing with other researchers, as needed. 

We counted any risk mentioned, regardless of detail or emphasis. This one-to-one mapping ensured comparability with repository statistics. In the end, we were able to compare the percentage of AI committee reports that mention a certain risk to the percentage of sources in the AI Risk Repository that mention the same risk.

\subsubsection{Thematic Analysis (RQ4).}\label{sss:thematic-analysis}
We chose a qualitative method to understand patterns in committee-recommended mitigation strategies for two reasons. First, no systematic framework existed to compare committee-proposed mitigation strategies to those documented in HCI literature. While an AI Mitigation Repository~\cite{ai-risk-mitigation-repository} exists, it is still under development, and incorporates policy documents beyond HCI, making direct comparison infeasible. Second, we wanted to explore deeper into the patterns we unearthed from the first three research questions. 
Two researchers met to collaboratively and used inductive thematic analysis~\cite{braun2006using} to identify overarching themes in committee reports. These themes were then refined through two rounds of iteration until consensus was reached. We lightly edited quotes presented in the findings for readability.

\section{Findings}

\begin{figure*}[!ht]
  \centering
  \includegraphics[width=0.75\linewidth]{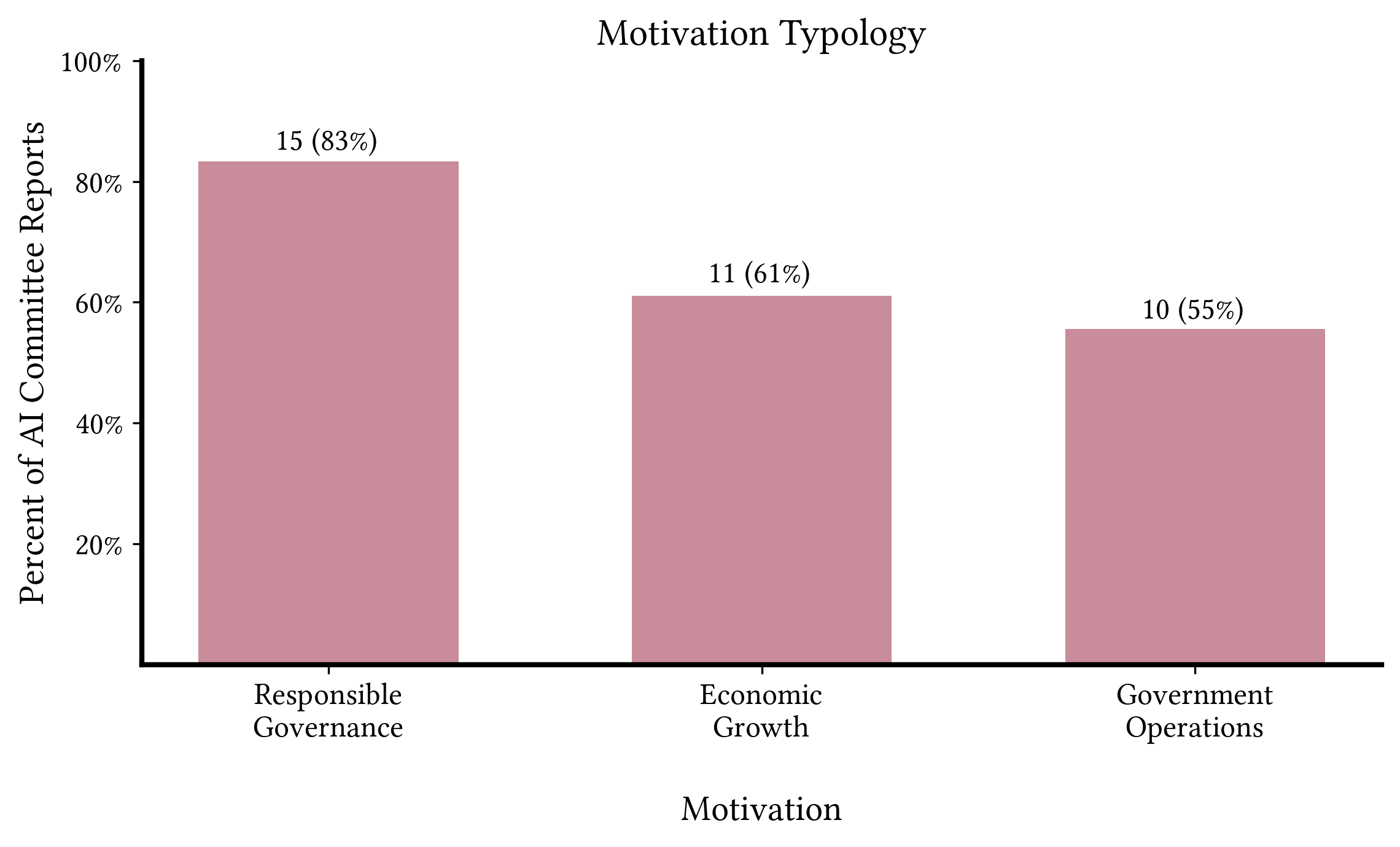}
  \caption{\textbf{Motivation Typology:} State-level policymakers were motivated to form AI committees to pursue responsible governance, promote economic growth, or enhance government services. State committees could express multiple motivations at once.}
  \Description{This bar chart shows the typology of state motivations for forming AI committees. The x-axis lists three categories: responsible governance, economic growth, and government operations. The y-axis shows the percentage of states citing each motivation: 15 states (83\%) for responsible governance, 11 (61\%) for economic growth, and 10 (55\%) for government operations.}
  \label{fig:motivation}
\end{figure*}

\subsection{Underlying Motivations (RQ1)}
The dominant factors driving the creation of AI committees centered on states' desires to leverage AI's benefits, specifically for improved governance and economic opportunities. State committees also overwhelmingly acknowledged that responsible governance was a necessary condition for realizing these advantages. The three factors, namely economic growith, government operations, and responsible governance, and their prevalence are summarized below (Figure \ref{fig:motivation}).

\subsubsection{Economic Growth.}
A critical motivation for states was to position themselves for economic competitiveness in an AI-driven economy. Eleven of the 18 state reports we analyzed sought to understand and identify how to integrate AI into their own economies and exploit current technological advances. States like Wisconsin exemplify this focus:

\begin{quote}
``Whereas establishing this task force will serve as a crucial mechanism to understand, adapt, and capitalize on the transformations generative artificial intelligence will bring.'' \textit{---WI}
\end{quote}

States discussed the importance of maintaining a competitive edge in the AI economy. They emphasized the need to foster innovation, with California investigating how ``Generative Artificial Intelligence (GenAI) can be responsibly used to spur innovation.'' Arkansas and Georgia shared this motivation, with Arkansas highlighting the need to remain "at the forefront of AI innovation."

Another focus for creating state committees was to upskill the workforce and prepare it for an AI-dominated economy. States like Arkansas aimed to do so ``by fostering an AI-ready workforce;'' meanwhile, Georgia sought to ``review the potential impacts of AI technology on the workforce'' and explore strategies for promoting workforce development. Similarly, California's mission was to prepare the workforce and create a welcoming business environment by ``lead[ing] in training and supporting workers, allowing them to participate in the AI economy and creating the demand for businesses to locate and hire here in California.''

\subsubsection{Government Operations.}
Ten reports highlighted how AI could enhance government operations, aiming to integrate AI into existing government services that improved service delivery or bolstered institutional capacity. For instance, the Georgia AI committee wrote that they ``[were] tasked with examining current and future uses of AI technologies in this state.''  Indiana similarly conducted a study of ``[how] artificial intelligence technology has been used, developed, or considered for use by state agencies'' in order to adopt it themselves.

However, emphasis on how exactly states should approach the adoption of AI differed. For some, the primary focus was on improving efficiency, effectiveness, or citizen experience. Arkansas, for instance, outlined its interest in streamlining operations and personalizing government services this way: 

\begin{table*}[!ht]
    \centering
    \caption{\textbf{Sector-Specific Benefits and Risks of AI.} Sectors identified by states as capable of being enhanced by AI, along with prioritized use-cases, while raising concerns about amplified or novel AI-related risks. We list our typology and provide examples from committee reports.}
    \Description{This table has four columns: (1) sector-specific domains, (2) their definitions, (3) examples of potential AI benefits, and (4) examples of potential AI risks. For instance, in the health sector, the definition is ``improve health outcomes and support patient care,'' with benefits such as early disease detection and risks such as incorrect diagnosis.}
    \label{tab:ai-applications}
    \begin{tabular}
    {@{} >{\raggedright\arraybackslash}p{2.2cm} >{\raggedright\arraybackslash}p{3.7cm} >
    {\raggedright\arraybackslash}p{4.0cm} >
    {\raggedright\arraybackslash}p{4.0cm} @{}}
     \toprule
        \textbf{Domain} & \textbf{Definition} & \textbf{AI Benefits} & \textbf{AI Risks} \\ \midrule
     
     \textbf{Cross-Sector} & & & \\ \midrule
      
      Government Services
      & Design, procurement, or deployment of AI systems to support government operations.
      & Chatbots for constituent support; legislative document drafting; public sentiment analysis. & Covert surveillance; handling sensitive data; threats to elections. \\ \midrule
     
      Workforce \& Economy
      & Support for workers, employers or businesses for better employment and economic outcomes.
      & Predict labor market trends; connect residents with tailored employment; reduce administrative burden. 
      & Worker displacement; decline of job quality; bias in hiring. \\ \midrule

    \textbf{Sector-Specific} & & &\\ \midrule
    Health
    & Improve health outcomes and support patient care.
    & Early disease detection; drug discovery; automation of patient record summaries.
    & Incorrect diagnosis; handling sensitive data; lack of human oversight.\\ \midrule
    Education
    & Enhance teaching and learning processes in K--12 and higher education.
    & AI-powered grading tools; virtual teaching assistants; standardized exam simulations. 
    & 
    Overreliance; handling sensitive data; educator job displacement.\\ \midrule
    Climate
    & Enhance environmental stewardship and sustainability.
    & Energy grid optimization; environmental monitoring; precision agriculture. 
    & High energy use; high water consumption.\\ \midrule
    
    Transportation
    & Improve mobility through better safety measures and transit efficiency.
    & Traffic monitoring; collision detection; autonomous vehicles.
    & Road safety with autonomous vehicles. \\ \midrule
    
    Public Safety
    & Support law enforcement, criminal justice, and rehabilitation systems. 
    & Tailored rehabilitation; criminal activity detection; post-incarceration employment matching.
    & Bias; handling sensitive data. \\ \midrule
    
    Entertainment \& Marketing
    & Enhance entertainment, marketing, and sales workflows. 
    & Generate social media content; website design; tailored recommendations. 
    & Intellectual property issues; displacement of jobs.\\ \midrule
   
   Agriculture
    & Modernize farming operations by improving productivity and reducing costs. 
    & Precision agriculture; crop yield improvement; resource efficiency. & (None mentioned.)\\
    \bottomrule
    \end{tabular}
\end{table*}

\begin{quote}
``[The goal is to] improve government services [and] leverage AI to enhance the efficiency, transparency, and effectiveness of state operations: (1)~Enhance citizen experiences and deliver more efficient and responsive government services. (2)~Reduce operational costs and improve resource allocation. (3)~Advance data-driven decision making across agencies.'' \textit{---AK}
\end{quote}

For others, such as Connecticut, strong emphasis was placed on \textit{ethical} government adoption, stating that ``the charge of the group was to make recommendations concerning ethical and equitable use of artificial intelligence by state government.'' Indiana similarly wanted to understand ``benefits and risks to state agencies of state agency use of artificial intelligence technology'' insofar as those might affect ``the rights and interests of Indiana residents.''

\subsubsection{Responsible Governance.}\label{sss:responsible-governance}
This motivation to ensure responsible AI adoption remained central across state AI committee reports and formed the most dominant mission-statement for state-level policymakers. Fifteen out of the 18 states consistently asserted that while AI presented significant opportunities for economic growth and improved public services, these benefits could be realized only ``when done safely and correctly,'' as Arkansas observed.

Most states adopted a nuanced approach, explicitly acknowledging both the benefits and inherent risks of AI systems. States consistently framed their policy objectives as striking an optimal balance between maximizing benefits and minimizing risks. Arkansas sought to ``outline a comprehensive strategy for Arkansas to maximize AI's benefits while maintaining strong protection for its residents,'' while California emphasized the need for ``coordinated and thoughtful public policies to mitigate risks and maintain public trust.''

Beyond risk-benefit calculations, some states prioritized establishing foundational governance frameworks. Oregon's committee, for example, focused on definitional clarity and examined ``terms and definitions related to artificial intelligence applied in technology-related fields that may be used for legislation.''

\subsection{Priorities of AI Benefits and Risks (RQ2)}

After defining motivations, states discussed AI’s benefits and risks within a largely sector-specific lens to assess its trade-offs. 
Additionally, since AI usage can span multiple sectors at once, committees also defined cross-sector applications and domains. We found that \emph{Government services} and \emph{Workforce \& economy} specifically were treated as such. 

Table \ref{tab:ai-applications} defines each sector and provides examples of how states envisioned AI enhancing or harming it. Sectors that were featured predominantly included government services, workforce and economy, followed by areas such as health, education, and climate, often tied to local strengths or concerns. Examples of AI-related benefits highlighted included the integration of chatbots constituents could use to interact with public services, support in legislative and regulatory document drafting, or analysis of public sentiment policy relevant issues. States further pointed to better employment matching opportunities, enhanced medical diagnoses, or personalized AI teaching assistants.

In terms of risks, committees cited concerns about data privacy, misuse, and leakage, along with overreliance and potential job displacement. What our findings reveal, however, is that \textit{AI-related risks were emphasized and discussed in far less depth than the associated benefits} for each sector, as seen in Figure~\ref{fig:benefit_v_risk}. Across states, discussions emphasized benefits over harms: for example, government services (30 (83\%) vs. 21 (58\%)), health (23 (63\%) vs. 17 (47\%)), and agriculture, where no risks were mentioned at all. 
For thoroughness, we compared the paired observations of benefits vs. risks across sectors and states using a Wilcoxon signed-rank test. The test produced a Wilcoxon statistic of 330.0 and a $\text{p-value} < 0.0001$, which was highly significant\footnote{We recognize that cluster and group-level effects in our data might affect the Wilcoxon signed-rank test. For example, benefit-risk score patterns may be similar within states. To address this, we conducted additional analysis for robustness detailed in Appendix~\ref{appendix:stattest}.}. 
In other words, the test confirms what is visually obvious: the distribution of risks was consistently lower than the distribution of benefits.

\subsection{Where AI Committee Reports and HCI Align and Misalign (RQ3)}\label{sec:RQ3}\label{sss:ai-risk}

In lieu of a sector-specific risk framing, states instead described risks in more general terms. They focused on bias, misinformation, overreliance, or transparency and unequal performance broadly. By contrast, \textit{risks prominent in literature from HCI and related fields (e.g., cultural devaluation of human effort) received little sustained attention in committee reports}.

\begin{table*}[!ht]
  \centering
  \caption{\textbf{AI Risks in Committee Reports vs HCI.} ``Proportion of HCI Literature'' is the share of HCI-related papers in the MIT AI Risk Repository mentioning the risk; ``Proportion of AI Committee Reports'' is the share of committee reports doing so. Pink with ${\dagger\dagger}$ marks risks mentioned significantly less often in reports than in the literature; dark green with ** marks those mentioned more. Pink with a single $\dagger$ indicates a weak, non-significant difference. Percentages for main categories (bold) are based on the share of papers/reports mentioning any subcategory.}
  \Description{This table has three columns: domain, proportion of HCI literature, and proportion of committee reports. “Proportion of HCI Literature” is the share of HCI-related papers in the AI Risk Repository mentioning the risk; “Proportion of AI Committee Reports” is the share of committee reports doing so. Color coding indicates differences: pink with two daggers for risks mentioned significantly less in reports, dark green with two asterisks for those mentioned more, and pink with one dagger for weak, non-significant differences. Results show that committee reports emphasize risks like unequal performance across groups and lack of transparency and accountability more than the HCI literature, but discuss lack of capability/robustness, economic and cultural devaluation of human effort, and loss of human agency less.}
  \label{tab:ai-risks}
  \begin{tabular}{@{} >{\raggedright\arraybackslash}p{9cm} >{\raggedleft\arraybackslash}p{2.5cm} >{\raggedleft\arraybackslash}p{2.5cm} @{}}
  \toprule \\
  Domain & Proportion of HCI Literature & Proportion of AI Committee Reports \\ \midrule
  \textbf{Discrimination \& Toxicity} & 70\% & 61\% \\ \midrule 
  Unfair discrimination and misrepresentation & 63\% & 61\% \\
  Exposure to toxic content & 33\% & 27\% \\
  Unequal performance across groups & 17\% & \textbf{\textcolor{darkfern}{50\%**}} \\ \midrule
  \textbf{Privacy \& Security} & 68\% & 44\% \\ \midrule
  Compromise of privacy by obtaining, leaking or correctly inferring sensitive information & 59\% & 44\% \\
  AI system security vulnerabilities and attacks & 37\% & 16\% \\ \midrule
  \textbf{Misinformation} & 46\% & 44\% \\ \midrule
  False or misleading information & 37\% & 44\% \\
  Pollution of information ecosystem and loss of consensus reality & 16\% & 6\% \\ \midrule
  \textbf{Malicious Actors \& Misuse} & 71\% & 61\% \\ \midrule
  Disinformation, surveillance, and influence at scale & 51\% & 50\% \\
  Cyberattacks, weapon development or use, and mass harm & 57\% & \textbf{\textcolor{darkMauve}{16\%$^{\dagger\dagger}$}} \\ 
  Fraud, scams, and targeted manipulation & 40\% & 44\% \\ \midrule
  \textbf{Human--Computer Interaction} & 49\% & 33\% \\ \midrule
  Overreliance and unsafe use & 32\% & 27\% \\
  Loss of human agency and autonomy & 33\% & \textbf{\textcolor{darkMauve}{5\%$^{\dagger\dagger}$}}  \\ \midrule
  \textbf{Socioeconomic \& Environmental} & 76\% & 61\% \\ \midrule
  Power centralization and unfair distribution of benefits & 41\% & 16\% \\ 
  Increased inequality and decline in employment quality & 41\% & 44\% \\
  Economic and cultural devaluation of human effort & 35\% & \textbf{\textcolor{darkMauve}{11\%$^{\dagger}$}} \\
  Competitive dynamics & 19\% & 5\% \\ 
  Governance failure & 30\% & 33\% \\
  Environmental harm & 38\% & 16\% \\ \midrule
  \textbf{AI System Safety, Failures, \& Limitations} & 75\% & 61\% \\ \midrule
  AI pursuing its own goals in conflict with human goals or values & 48\% & \textbf{\textcolor{darkMauve}{0\%$^{\dagger\dagger}$}} \\
  AI possessing dangerous capabilities & 25\% & 11\% \\ 
  Lack of capability or robustness & 56\% & \textbf{\textcolor{darkMauve}{22\%$^{\dagger}$}}  \\
  Lack of transparency or interpretability & 33\% & \textbf{\textcolor{darkfern}{50\%**}} \\ 
  AI welfare and rights & 3\% & 0\% \\
  Multi-agent risk & 3\% & 0\% \\
 \bottomrule
  \end{tabular}
\end{table*}

Table \ref{tab:ai-risks} compares the frequency specific risks appeared from the AI Risk Repository's sample of HCI and related papers to the frequency that those risks appeared in state committee reports. While these frequencies are only a rough proxy of attention to a topic, they can nevertheless tell us how many states find a particular AI risk important to discuss. We ran a chi-squared test across risk subdomains on counts of committee reports vs.\ HCI literature. The result ($\chi^2 = 40.04$, $p = 0.015$) showed a significant difference in risk distributions between the two sources. Put differently, \textit{risks discussed in committee reports differ significantly from those discussed in HCI.}

To identify which subdomains drove this difference, we examined standardized residuals (using $|\text{residual}| \geq$ 1.28, or 90\% confidence) as the cutoff for indicating whether a risk was over- or under-represented in committee reports relative to HCI literature. We found overlap in which common risks were discussed. For example, 70\% of the HCI literature and 61\% of committee reports highlight discrimination and toxicity. Similarly, the risk of misinformation was discussed in 46\% of relevant HCI papers and 44\% of AI committee reports. State reports also stressed risks associated with human--computer interactions, such as overreliance in decision-making: 32\% of the relevant HCI literature mention this, versus 27\% of the committee reports.

However, we also found key areas of divergence. Unlike HCI literature (48\%), none of the committee reports gave attention to risks related to conflict with human goals or values or AI pursuing its own goals in. Only 13\% of the committee reports mentioned the risk of cyberattacks, weapon development or use, and mass harm, despite this being a prominently discussed concern in the HCI literature with 57\% of papers mentioning it. In short, \textit{high-risk, low-likelihood scenarios were significantly underemphasized} in committee reports compared to HCI literature based on statistical analysis of standardized residuals.

Reports placed much stronger emphasis on unequal performance across groups (50\% vs. 17\%), often attributing these harms to biased training data that produced unequal outcomes. They also disproportionately highlighted lack of transparency and interpretability (50\% vs. 33\%), frequently framing these as gateway risks that enabled unfair systems or overreliance. The high co-occurrence of these AI risks are linked; many reports argued that greater transparency could help mitigate issues such as algorithmic bias.

\subsection{Mitigation Strategies for AI Risks (RQ4)}\label{ss:mitigation} 
Committees proposed a slew of mitigation strategies. These were often presented separate from associated risks. We first detail recommendations prominent within committee reports (Section~\ref{sss:mitigation}). We then situate the recommendations within their greater contexts (Section~\ref{sss:greater-context}). 

\subsubsection{Mitigation Strategies.}\label{sss:mitigation}

\paragraph{\textbf{AI Literacy.}}\label{sss:mitigation-literacy}
A predominant mitigation strategy across state committee reports centered on improving AI literacy for K--12 students, the general public, and the workforce. California sought to ``explore specialized training and development curricula,'' while Vermont recommended creating ``programs to increase awareness of artificial intelligence among students, teachers and the general public.'' Similarly, New Jersey directed agencies to ``promote digital literacy, engagement, and increased exposure to AI tools and resources.''

States proposed varied approaches to building AI literacy. Connecticut suggested policymakers create a ``citizens AI academy [and] curat[e] online courses.'' Wisconsin described its proposed ``AI workforce talent pipeline'' as preparing residents and ``help[ing] raise awareness of, allow career exploration in, and provide training and worker upskilling in AI-related occupations.'' Illinois proposed allocating ``funding and grants to research the impact of GenAI on education and its potential to enhance learning outcomes.''

\paragraph{\textbf{Risk Assessments.}}\label{sss:mitigation-risk-assessment}
Beyond AI literacy and upskilling, committees also examined AI systems themselves. They repeatedly emphasized the need for pre-deployment and ongoing risk assessment procedures. Texas proposed to ``mandate risk assessments for high-risk AI deployment by public and private actors,'' while Illinois recommended ``initial and regular risk assessments to identify risks\ldots throughout [the AI] lifecycle.''

\paragraph{\textbf{Inclusive Governance.}}\label{sss:public-inclusion}
Another frequently mentioned mitigation strategy was inclusive and diverse governance. States stressed that diverse stakeholder input was crucial. Oklahoma’s committee made this explicit, stating that ``deployment of [AI] should involve public input and oversight to confirm that AI serves the public interest and upholds democratic values.'' Connecticut proposed creating a ``permanent advisory committee composed of representatives from industry, academic, and government,'' while Wisconsin argued that workers should ``have a seat at the table'' as decisions are made about AI. 
States further emphasized the need to expand public participation and diversity, not only during the development and deployment of AI systems, but during the formation of AI committees, writing of reports, and policymaking themselves. An exemplar of this call, Washington's committee report urged the state to ``broaden [public] engagement in these conversations.''

\paragraph{\textbf{Human-Centered Values.}}\label{sss:mitigation-values}
AI committee reports recommended that policymakers embrace values such as privacy and security, human oversight, and transparency. Alaska urged the state to ``prioritize privacy,'' suggesting that ``AI systems be designed and deployed in a manner that protects sensitive and regulated data.'' Illinois positioned data privacy as a core right, insisting that ``policymakers take measures to give consumers control over their personal data.''

States also recommended human oversight as an essential safeguard, deeming human-in-the-loop systems imperative. Georgia recommended that policymakers ``encourage the adoption of human-in-the-loop and human-on-the-loop frameworks for AI systems, particularly in sensitive sectors,'' and New Jersey directed agencies to ``ensure that humans remain actively involved when GenAI tools are deployed.'' Washington similarly stressed the need for humans to ``retain appropriate human agency and oversight.''

Another mitigation category focused on transparency and open communication. Connecticut's ``final recommendations center[ed] around how to promote transparency and accountability,'' and Texas proposed to ``implement transparency standards for high-risk AI in the private sector.''

In practice, state committees suggested achieving these values through developing new frameworks or adopting existing ones. New Jersey proposed that policymakers ``develop an ethical use framework\ldots  includ[ing] recommendations to protect student data... emphasize transparency and understanding of potential biases and inaccuracies in AI outputs.'' Oregon's committee similarly stressed the importance of establishing ``a framework to determine use cases of AI that will require `human in the loop.' ''

\subsubsection{Mitigation Strategies in Context.}\label{sss:greater-context}

\paragraph{\textbf{Ambiguity in Committee Recommendations.}}\label{sss:definitional-ambiguity}
While reports echoed HCI considerations, promoting AI literacy, diverse inclusion, and human-centered design values,
%
\textit{many recommendations remained under-specified}, often invoking terms such as ``high-risk,'' ``misuse,'' or ``transparency'' without identifying their operational meaning. Several committees acknowledged this lack of definitional precision, with Colorado noting a need for ``more specifically defining the types of decisions that qualify as `consequential decisions' under the law.''

Definitional ambiguities were widespread. Most importantly, committees struggled to define AI itself, the very subject of their inquiry. Some reports offered no definition, while Oregon's committee indicated that producing one was its very purpose.

Several states distinguished between AI broadly and generative AI (GenAI) specifically. Definitions of GenAI emphasized output characteristics---``systems that simulate human-produced content'' (IL)---rather than underlying mechanisms. New Jersey similarly defined GenAI as producing ``new content, such as text, images, and videos, based on simple user inputs.''

This definitional ambiguity frequently resulted in \textit{recommendations that called primarily for more deliberation, more data collection, or more committees}. Vermont explicitly concluded that it ``does not recommend the promulgation of new, specific State regulations of artificial intelligence at this time,'' while Connecticut recommended ``establishing a permanent advisory committee.'' Georgia urged the legislature to ``continue statewide efforts to monitor and update state law'' without explaining which laws to update or how. This \textit{under-specification resulted in generic recommendations}, such as Alaska's mitigation strategy to simply ``enact legal protections against misuse of AI.''

The lack of cohesion in defining AI further points to \textit{knowledge gaps whereby committee members lacked specificity in differentiating and defining AI systems}. To address these distinctions, \textit{committee reports frequently drew analogies to other technological eras} to contextualize AI. Illinois compared GenAI's transformative potential to ``electricity and the internal combustion engine,'' while Massachusetts likened it to the space race, noting that ``space science remains relevant today as we consider the implications of [AI].''

\paragraph{\textbf{Limited Diversity in Committee Participation.}}\label{sss:performance-public-inclusion}
Similar ambiguities plagued committee mitigation strategies that called for inclusive governance. \textit{Diversity was generally limited in practice}. On the one hand, states emphasized the importance of gathering diverse input, sometimes recruiting expert panelists to present to committee members. Connecticut, for instance, described hosting seven meetings with more than 20 speakers ``from around the world who all possessed some expertise regarding an aspect of artificial intelligence and artificial intelligence regulations.'' Nonetheless, though committees featured robust participation from industry and government officials, fewer panels included labor organizations or community groups, raising questions about whose perspectives were most influential in shaping policy recommendations.

As a case in point, New Jersey highlighted that they ``should prioritize underserved communities.'' They described how inclusivity is crucial for forming fair policy recommendations: ``The State should promote public engagement and inclusivity in AI to broaden access to AI technology and to ensure that it benefits all segments of society.'' Furthermore, New Jersey's committee emphasized that its ``contributors to the creation of this report [were] members of State Government, Academia, and the Private Sector,'' suggesting broad diversity.

In practice, however, representation was skewed: of the 35 external participants who informed the committee, a clear majority (19) came from industry, including Microsoft, startups, and other for-profit organizations. Academia accounted for 11 representatives, while civic groups were represented by only three individuals, one of whom was affiliated with a governmental organization and another with a non-profit. 
Indeed, \textit{industry, followed by government, and academic stakeholders appear to have played a more central role} in shaping the deliberations and recommendations of most state AI committees. Kentucky, similarly, for instance, noted that its committee heard from government representatives as well as individuals from Microsoft and Google.

The stakeholders committees involved were intertwined with the idea that committee reports consistently cast AI as transformative and unavoidable. Oklahoma portrayed AI as having ``the potential to revolutionize the way our society operates.'' Colorado discussed how this is the ``early [stage] in this latest technological revolution,'' while Massachusets asserted that ``we are now at a real inflection point.'' This framing positions the present moment as a critical juncture during which states have to act. 

With an AI-future looming, states stressed the urgency of immediate action. Vermont advocated for ``immediate investment in this area'' to ''maximize potential economic benefits and help keep Vermont at the forefront of this technological revolution.'' New Jersey further noted that:

\begin{quote}
``AI technologies are evolving rapidly, and some may believe it is best to wait until GenAI is better understood before adopting new GenAI models. However, the learning curve is too steep to allow delay. The longer the State takes to expand adoption, the harder it will become.'' \textit{---NJ}
\end{quote}

AI was thus seen as an unstoppable force rather than a technology subject to deliberation. The reports assumed AI's integration into society was inevitable and required rapid action, yet this fast-paced rhetoric conflicted with principles of inclusive and diverse governance.

\section{Discussion}
States articulate multiple motivations for forming AI committees. Our findings show that \textbf{AI committees are motivated to enact responsible governance}. Throughout their tenure, many additionally recommended mitigation strategies that align with HCI's human-centered values and promoted data privacy policies, transparency in systems, and risk assessments. However, we also found that much of \textbf{this engagement was cursory, superficial, and lacked specificity}. Across the board, state reports emphasized benefits over potential AI-related risks within economic sectors. The mitigation strategies that states proposed were further full of definitional ambiguities, leading to an underspecified and non-operationalized understanding of next steps.

On the one hand, it is heartening that states and HCI align on human-centric AI and ethical values; on the other, \textbf{several concerns discussed by HCI scholars are under-explored in U.S.\ state policymaking}. Ongoing work at the intersection of HCI and policy is vital~\cite{feng2025sociotechnical, bowman2022measuring, reuel2024position} as policies are actively being set and HCI methods can support filling relevant gaps. HCI scholarship offers tools for understanding how people interact with AI and for using participatory methods to include a diversity of voices. In the following subsections, we contextualize our findings within existing HCI literature (Sections~\ref{ss:risk-regulation} to \ref{ss:gaps-in-sociotechnical}) and expand on where key differences lie. Finally, we highlight how HCI approaches can help support future regulatory efforts (Section~\ref{ss:agenda}).

\subsection{Risk Regulation as Instrumental Approach}\label{ss:risk-regulation}
Policymakers consistently framed AI as a transformative technology (Section~\ref{sss:performance-public-inclusion}) that brings significant benefits and whose posed dangers can be remediated through regulation (Section~\ref{sss:responsible-governance}). This echoed tenants of Kaminski et al.'s ``ex ante risk regulation''~\cite{kaminski2023regulating}. While we did analyze policy reports from a risk-benefit perspective ourselves, playing into the ex ante framework, our findings notheless highlight the specific nature in which committee reports play into this framing. 

First, reports frequently presented AI as both unavoidable and revolutionary, positioning governance as a matter for risk mitigation only, rather than of critical scrutiny. This stance is further evidenced by the fact that policy discussions emerged \textit{after} widespread deployment of AI technologies~\cite{wyatt2023technological}, reflecting a presumptive approach where innovation leads and regulation follows. Moreover, industry voices carried disproportionate weight in contributing to committees, and the possibility of slowing or refusing AI adoption was largely absent.
Second, reports framed harms as individual rather than systemic. For example, when addressing algorithmic bias, committees reduced the problem to requiring model risk assessments. This narrow lens elided broader structural issues: How should risks and biases be defined? What if models cannot be improved because the necessary data does not exist, or because the underlying systems, e.g., policing, finance, housing, are the result of deeply rooted inequalities?

Researchers have noted that the notion of risk is ``a \textit{design decision} as much as any other,''~\cite{luusua2020artificial} and, instead of being objective, it is rather shaped by social and cultural values and perceptions.

Thus, scholars are pointing to alternative ways of approaching AI regulation beyond simply risk mitigation. One option is the \textit{precautionary principle}~\cite{kriebel2001precautionary}, which postulates that rather than asking how harms can be managed once a technology is adopted, we should shift the burden of proof to those promoting the technology. In this framing, the central question is not around how we minimize risks, but rather what evidence of benefits exist to justify the proliferation?

\subsection{Gaps in Diverse Representation}\label{ss:gaps-in-representation}
As Social Construction of Technology theory~\cite{douglas2012social, mahdavi2024social} and reconstructionist scholars indicate, the role that a technology occupies within society is not pre-determined, but rather emerges through an active process of push and pull and the social construction of its meaning. 
Although this theory does not prescribe who should shape technologies, it helps illuminate how a diverse set of groups can hold competing understandings that influence the trajectory of technology development and governance. 

HCI researchers in that vein have shown that the perceptions of AI from individuals in underrepresented groups, such as women and people of color, in particular, often differ significantly from those of the general public~\cite{moreira2025hall}. 
Consequently, it is necessary for a broad range of perspectives to be involved in the social construction of AI technologies. These perspectives include civil society organizations, labor unions, community advocates, and affected populations, and so forth, to define what a technology is, what problems it addresses, and how it ought to be governed. 

We found that, while public participation and diverse involvement were nominally encouraged in committee reports (Section~\ref{sss:performance-public-inclusion}), in practice, the creation of policy documents often excluded many important stakeholders (Section~\ref{sss:performance-public-inclusion})~\cite{young2019toward, kurath2009informing, guston1999evaluating}. Schiff et al.~\cite{schiff2020s} cautioned that such processes risk becoming primarily performative, signaling social responsibility rather than embodying it.

\subsection{Diminished Socio-Technical Perspectives}\label{ss:gaps-in-sociotechnical}
The risk regulation orientation and lack of diverse stakeholder representation reflect a deeper, more fundamental lack in AI policymaking. Whereas HCI researchers have long emphasized issues such as the centralization of power~\cite{kalluri_2024}, the unequal distribution of benefits, the devaluation of human labor~\cite{sarkar2023enough, woodruff2024knowledge}, and the erosion of human agency and autonomy~\cite{meek2016managing, shelby2023sociotechnical}, these themes were largely absent from committee reports. Instead, policymakers tended to frame risks in narrower, more technical terms, leaving unexamined the broader social and cultural dimensions in which AI technologies are embedded.

This lack of contextualization overlooked global labor infrastructures that underpin AI development, including the dependence on low-wage data workers in the Global South~\cite{Perrigo_2023a} who label and curate massive datasets. By failing to acknowledge these hidden forms of labor, \textbf{the reports reinforced the impression that AI is a clean, automated, and inevitable innovation, not a technology built on human effort and global inequities}.

The same pattern appeared in discussions of AI’s environmental impact. Reports often highlighted the potential benefits of AI for addressing climate and sustainability goals but scarcely mentioned the vast  environmental costs of developing and deploying these systems. The enormous energy demands of training large-scale models~\cite{falk2025carboncradletograveenvironmentalimpacts, luccioni2025efficiency, 10.1145/3715275.3732006}, for instance, were rarely documented, even as such demands threaten to undermine already fragile climate commitments across both government and industry~\cite{Roytburg_2024, WITHROW_2025}. These omissions are symptomatic of current trends in AI governance, where ``technical considerations push aside an accounting of social factors''~\cite{solow2023can}. The reports we reviewed similarly blurred this distinction, treating capabilities and benefits as interchangeable and failing to note the 
assumptions imbued in their conceptualization.

\subsection{Future Work}\label{ss:agenda}
HCI’s foundational commitment to situating technology within social contexts provides critical opportunities for more comprehensive approaches to AI policy~\cite{dym2022building, jackson2014policy}. Generally, HCI's socio-technical lens envisions policymakers as architects of technological change rather than bystanders, and we believe that this lens can lend support to ongoing regulatory efforts. HCI methods can broaden the scope of who is involved in governance efforts, and what those efforts entail.

\subsubsection{Diverse Stakeholder Input and AI Literacy.} 
State committee reports voice support for public participation and the democratization of AI and its governance. However, based on a careful reading of these reports, such commitments remain superficial and are not integrated into action plans for how the committees will move forward. HCI’s work on citizen science can support real-time collection of public perceptions. For example, HCI researchers have built tools that can capture public opinion about AI~\cite{moreira2025hall}; these tools could be modified and applied to include information on current policy initiatives and enable public debate. Since many feel un- or under-informed about AI~\cite{xie2025exploring}, promoting AI literacy can empower broader participation and help individuals distinguish between AI’s capabilities, how these could translate into benefits, and the limitations and risks AI can impose. 

Furthermore, HCI could employ participatory design methods, such as Diverse Voices~\cite{young2019toward}, which draws from Value Sensitive Design. This methodology describes how perspectives of underrepresented groups can be centered when drafting technology policy documents. Similarly, Krafft et al.~\cite{krafft2021action} frame AI as a public policy issue, supporting grassroots efforts to navigate local ordinances on AI and surveillance; their 
Algorithmic Equity Toolkit~\cite{krafft2021action} offers practical tools for public engagement, including a working definition of AI, methods for analyzing AI systems, and questions to guide advocacy directed at policymakers and agencies. HCI researchers could utilize similar approaches to help translate abstract notions of AI risks into more specific concepts that could guide development of rules and regulations~\cite{davis2012occupy}. Through these efforts, its important to keep in mind that participatory AI does not replace democratic processes, nor should it be required to take on legislative functions~\cite{10.1145/3551624.3555290}.

\subsubsection{Standardizing Definitional Practices.} 
Across committee reports, definitions of AI varied considerably. Some reports omitted definitions altogether, and only a few distinguished between AI and GenAI. Provided definitions rarely offered the kind of technical or legal rigor necessary for policy implementation. This pattern was mentioned by Maas~\cite{maas2023concepts}, who highlights persistent terminological ambiguities in AI governance at large.
Importantly, the language used to describe AI heavily influences the scope and enforceability of legal and policy regimes~\cite{maas2023concepts, maas2023ai}, and many researchers emphasize that specificity is a prerequisite for effective policy development~\cite{schiff2020s, maas2023concepts}. This contrasts with areas where AI definitions have already become central to governance efforts, with ``general-purpose AI'' codified in the EU AI Act~\cite{benifei2023proposal} and ``frontier AI models'' utilized during the 2023 UK AI Safety Summit~\cite{gov_uk}. 

Instead of definitional precision, many state committee reports relied on inconsistent metaphors to contextualize AI. Some states compared AI to the internet, others to the industrial revolution, and others still to the space race. These metaphors may serve as helpful anchors for policymakers, shaping how they interpret a technology’s impact and whether it requires new regulatory frameworks~\cite{wong2015wireless}. If AI were framed as akin to the internet, regulation would emphasize interoperability, open-access standards, and issues such as consumer rights and data protection, much like net neutrality debates. In contrast, if it were cast as a space race, regulations would prioritize government investment in research and development and focus on national security and military applications.
For HCI, expanding the use of metaphors as design tools for communicating policy implications could support the creation of more cohesive and standardized definitions. More deliberate choices of metaphors could make the stakes of AI governance more comprehensible and reduce inconsistencies~\cite{yang2024future}.

\subsubsection{HCI--Policy Bridge.}
Significant ideological differences remain between HCI and policymakers, stemming not only from language scope but from broader challenges of collaboration~\cite{yang2024future}. Questions such as the laws and policies HCI should prioritize in its research remain unresolved, and differences in priorities, perceptions of risk, and timelines, compounded by institutional barriers, further widen the gap~\cite{yang2024future, spaa2019understanding}.

HCI scholarship could strengthen its policy relevance by more explicitly addressing the scope and generalizability of its findings, asking questions such as: Is a technological harm specific to a context or population? Should it be addressed by global, national, local, or sector-specific  regulation? Just as HCI researchers have built toolkits to help community advocates navigate AI policy~\cite{krafft2021action}, similar design processes could guide researchers in crafting meaningful policy implications. A framework for contextualizing findings and linking them to policy contributions would make research more accessible to policymakers~\cite{yang2024future}.
Still, we realize that the work of researchers themselves reflects normative choices made within political and economic constraints.

\subsection{Limitations}
Our analysis has limits in generalizability. We focused only on state-level committee reports, introducing selection bias since these states had already taken initiative on AI. We also examined committee reports rather than legislation since legislative texts reveal little about deliberative reasoning and mostly capture only those issues salient enough for formal lawmaking. Further, recordings of deliberations are not always public~\cite{NCSL_2025b}, and we did not conduct interviews given the difficulty of recruiting state officials, though doing so could have yielded richer insights. In addition, our portrayal of policymakers’ perspectives relies on proxies; without interviews, their personal views remain unknown.
Finally, our analysis is U.S.-centric, focusing only on reports from U.S. policymakers. We made this choice given the country’s predominant role in AI investment and deployment, which shapes global trends. However, significant regulatory initiatives are emerging worldwide and warrant attention.

\section{Conclusion}
AI policy in the U.S.\ is actively taking shape. In the absence of comprehensive federal guidance, state governments are now forming AI committees, whose reports offer rare insight into how policymakers are interpreting AI and its trade-offs. To understand these dynamics, we analyzed 18 state-level committee reports and compared them to the AI Risk Repository. Our findings reveal both overlap and divergence between policymakers’ concerns and those of HCI scholars. Though states on paper noted the importance of responsible AI governance, their reports showed a lacked a critical socio-technical perspective. By interpreting these reports' findings through the lens of HCI research, we show how HCI methods can support policymaking in this formative period.
\section*{Acknowledgements}
We are grateful to the anonymous reviewers for all their feedback that helped improve the paper. Likewise, we want to thank the the members of the Wildlab\footnote{https://wildlab.cs.washington.edu/} who provided thoughtful insights and suggestions. This work was supported by the National Science Foundation Grant \#2315937. The findings and recommendations discussed in this paper are those of the authors and not necessarily those of the supporting bodies.

\bibliographystyle{ACM-Reference-Format}
\bibliography{sample-base}

\appendix

\section{Mixed-Effects Logistic Regression}\label{appendix:stattest}

We conducted a paired Wilcoxon signed-rank test comparing paired benefit and risk scores across states and sectors. Although the Wilcoxon test provides a nonparametric assessment of paired differences, however, it does not incorporate clustering effects that might have arisen due to state- or sector-level similarities. Because observations may not be independent within these groups, we also fit a logistic regression models with random effects for each grouping factor for robustness.

We modeled the binary outcome \texttt{Condition} (coded 0/1 for risk and benefit respectively) as a function of the ordinal predictor \texttt{Score}. We ran a logistic model rather than a linear regression of the form \texttt{Score}~\textasciitilde~\texttt{Condition} because the ordinal scale of the score variable would have violated linear regression assumptions. Since we are not interested in causal factors, but only in whether there are significant differences between benefit and risk scores, predicting the binary condition given the ordinal score was deemed appropriate.

For each grouping factor $g \in \{\texttt{State}, \texttt{Sector}, \texttt{State:Sector}\}$ we estimate the mixed model:
\[
\text{logit}\big(P(\texttt{Condition}_i = 1)\big)
    = \beta_0 + \beta_1 \,\texttt{Score}_i + u_{g(i)},
\]
where $u_{g(i)} \sim \mathcal{N}(0,\sigma^2_g)$ is a random intercept capturing group-specific baseline differences.

Across all models, the coefficient for \texttt{Score} was positive, stable, and highly statistically significant (all $p < 0.001$). The estimated odds ratio associated with a one-unit increase in \texttt{Score} was approximately 1.684, and the 95\% confidence interval was [1.35, 2.11]. Table~\ref{tab:score_effects} summarizes the fixed-effects estimates for these three specifications.

Our results indicate that the \texttt{Score} variable strongly predicted the binary variable \texttt{Condition}, referring to either benefits discussed within a report (\texttt{Condition} = 1) or risks (\texttt{Condition} = 0). This remained consistent even after accounting for group-level heterogeneity. Importantly, this consistency revealed that the effect we saw in the Wilcoxon-signed rank test was not driven by clustering effects across groups and reinforced our previous conclusion: higher 
\texttt{Score} values are systematically associated with benefits, and lower \texttt{Score} values are systematically associated with risks.

\begin{table*}[!ht]
\caption{Fixed-effect estimates for the \texttt{Score} in a mixed-effects logistic regression models.}
\centering
\begin{tabular}{@{} lllll @{}}
\toprule
\textbf{Model} & \textbf{\texttt{Score} Coefficient} & \textbf{Std. Error }&\textbf{ $z$-value} & \textbf{$p$-value} \\
\midrule
\texttt{Condition} $\sim$ \texttt{Score + (1|State)} & 0.521 & 0.114 & 4.57 & $<0.001$ \\
\texttt{Condition} $\sim$ \texttt{Score + (1|Sector)} & 0.521 & 0.106 & 4.91 & $<0.001$ \\
\texttt{Condition} $\sim$ \texttt{Score + (1|State:Sector)} & 0.521 & 0.095 & 5.50 & $<0.001$ \\
\bottomrule
\end{tabular}
\label{tab:score_effects}
\end{table*}

\section{Codebook}\label{appendix:codebook}
See Table~\ref{tab:codebook}.

\begin{table*}[!ht]
\centering
\caption{Codebook. \label{tab:codebook}}
\begin{tabular}{@{} >{\raggedright\arraybackslash}p{2.4cm} >{\raggedright\arraybackslash}p{5cm} >{\raggedright\arraybackslash}p{8.5cm} @{}}

\toprule 
\textbf{High-Level Code} & \textbf{Low-Level Code} & \textbf{Definition} \\

\midrule

Motivation & Lack of Federal Oversight & Absence or insufficiency of regulatory frameworks, guidelines, or enforcement mechanisms at the federal level to monitor, control, or govern AI. \\

& Unintended Consequences Concerns & Worry about the potential negative outcomes or side effects that may arise from the development, deployment, or use of AI. \\

& Expand AI and Lead AI Innovation & Compete with other states to be at the forefront of technology and innovation. \\

& Harnessing Benefits of AI & Leverage positive aspects and capabilities of AI for improved efficiency, innovation or better decision-making. \\

& Understand \& Prepare for Societal Impacts & Desire to study the impact of AI on various social, economic, environmental, etc. dimensions (positive or negative) or prepare citizens for the coming AI transition. \\

\midrule

Focus Area & Education & Understanding opportunities and challenges for AI in education at large, including how AI is used, or how AI literacy, etc. \\

& Government Operations & Understanding opportunities and challenges for AI within government, including potential uses of AI by state agencies, guidelines for public sector procurement, use, and monitoring of AI services, training for state employees on AI, assessing the impact of AI on state security, etc. \\

& Climate \& Energy & Understanding climate impact of AI systems, including the impacts of the increased need for energy to develop and deploy AI systems, or ability of public and private entities to increase energy efficiency and meet climate goals, etc. \\

& Entertainment & Understanding opportunities and challenges for AI in the entertainment industry, including unauthorized use of copyrighted or proprietary material, etc. \\

& Healthcare \& Accessibility & Understanding opportunities and challenges for AI healthcare, including assisting in medical decisions-making (e.g., diagnostics or treatment recommendations), involvement in activities related to patient care (e.g., email communications, visit summaries), etc. \\

& Labor/Employment & Understanding how AI impacts workers and employment, including worker protections and benefits, job displacement, job quality and worker well-being, education and reskilling, etc. \\

& Transportation & Understanding opportunities and challenges of AI used in the transportation sector, including optimizing transit routes, proliferating autonomous vehicles, etc. \\

& Public Safety & Understanding opportunities and challenges of AI used in the public safety, including criminal activity detection, post-incarceration service matching, etc. \\

& Agriculture & Understanding opportunities and challenges of AI in the agricultural sector, better water management, crop yield, etc. \\

\midrule

Mitigation Strategies & Create Future Committees & Ensure continued discussion among stakeholders by forming or continuing AI committees. \\

\bottomrule
\end{tabular}
\label{table:codebook}
\end{table*}

\begin{table*}[!ht]
\centering
\begin{tabular}{@{} >{\raggedright\arraybackslash}p{2.4cm} >{\raggedright\arraybackslash}p{5cm} >{\raggedright\arraybackslash}p{8.5cm} @{}}

\toprule 
\textbf{High-Level Code} & \textbf{Low-Level Code} & \textbf{Definition} \\ \midrule

Mitigation Strategies & Human-In-The-Loop & Prevent exclusive existence of automated decision-making by ensuring there is a human decision-maker in the loop with authority to challenge, review and overturn AI recommendations. \\

& Clearer Specifications & Create clearer and more legally relevant decisions for key terms, such as ``AI''; or better specify exceptions of created policies. \\

& Diverse Inclusion & Work with various stakeholders, in collaboration to design policies or technology governance frameworks. This can include working with underserved communities for input in the design, development, testing, training and oversight of technologies, working with universities and institutions to create an AI education program, etc. \\

& Funding Opportunities or Incentives & Incentivise AI use through funding or other opportunities, including funding to develop AI for different sectors, providing tax incentives for AI startups, etc. \\

& Improve AI Literacy & Encourage training and AI literacy of diverse groups. This can be done through organizing professional development workshops, providing resources or guidelines, etc. \\

& Maintain Data Privacy & Ensure proper data privacy guidelines are followed, such as anonymizing the data whenever possible.\\

& Transparency & Promote transparency of AI when used in a system including for job application screenings, employment, AI use within public services, etc. \\
\bottomrule
\end{tabular}
\label{table:codebook2}
\end{table*}

\end{document}